\documentstyle[12pt,aaspp4]{article}
\begin{document}
\title{Observation of the Halo of NGC~3077 Near the ``Garland'' Region
Using the Hubble Space Telescope$^{1}$}

\altaffiltext{1}{Based on observations with the NASA/ESA 
{\it Hubble Space Telescope}, obtained at the Space Telescope Science Institute, 
operated by AURA, Inc. under NASA contract No. NAS5-26555.}

\begin{center}
Shoko Sakai$^{2,3}$  and Barry F. Madore$^{4,5}$
\end{center}

\altaffiltext{2}{Kitt Peak National Observatory, National Optical Astronomy Observatories, Tucson 
AZ 85726}

\altaffiltext{3}{Division of Astronomy and Astrophysics, University of California, Los Angeles,
Los Angeles, CA 90095}

\altaffiltext{4}{NASA/IPAC Extragalactic Database, 
	California Institute of Technology, Pasadena CA 91125}
\altaffiltext{5}{Carnegie Observatories, Pasadena CA 91101}

\centerline{Running Headline: {\it NGC 3077}}

\def\Deg{\hbox{${}^\circ$\llap{.}}}
\def\Min{\hbox{${}^{\prime}$\llap{.}}}
\def\Sec{\hbox{${}^{\prime\prime}$\llap{.}}}

\def\ITRGB{$I_{\mbox{\tiny TRGB}}$}

\begin{abstract}

We report the detection of upper main sequence stars and red giant branch
stars in the halo of an amorphous galaxy, NGC~3077.
The observations were made using Wide Field Planetary Camera~2 on board the
Hubble Space Telescope.
The red giant branch luminosity function in I-band shows a sudden discontinuity
at $I = 24.0 \pm 0.1$ mag.
Identifying this with the tip of the red giant branch (TRGB), and adopting
the calibration provided by Lee, Freedman, \& Madore (1993) and the foreground
extinction of $A_B = 0.21$ mag,  we obtain a distance modulus of 
$(m-M)_0 = 27.93 \pm 0.14_r \pm 0.16_s$.  
This value agrees well with the distance estimates of four other galaxies
in the M81 Group.
In addition to the RGB stars, we observe a concentration of upper main sequence
stars in the halo of NGC~3077, which coincides partially with a feature known as
the ``Garland''.  Using Padua isochrones, these stars are estimated to be
$<$150 Myrs old.  Assuming that the nearest encounter between NGC~3077 and M81
occurred 280 Myrs ago as predicted by the numerical simulations (Yun 1997),
the observed upper main sequence stars are likely the results of the star
formation triggered by the M81-NGC~3077 tidal interaction.

\end{abstract}

\section{Introduction}

Located in a small group of galaxies around M81, NGC~3077
is a low-luminosity peculiar galaxy classified as Irr II (Sandage 1961),
or I0 pec (de Vaucouleurs, de Vaucouleurs \& Corwin 1976).
It was once suggested that this galaxy was a Seyfert galaxy 
(Wisniewski \& Kleinmann 1968), but Demoulin (1969) later noted that
its emission lines were too narrow to put it in the Seyfert class.
In the optical, NGC~3077 is characterized by its dusty appearance (Sandage 1961, Barbieri, Bertola \&
di Tullio 1974: hereafter BBdT74), many compact blue knots (BBdT74),  and 
its very high $M_H/L_B$ for typical S0 or I0 galaxies.
Its metallicity is also higher by about a factor of two 
than expected for a small galaxy (Heckman 1980).
Based on the spectral distribution covering 1000 -- 9000\AA,
Benacchio \& Galletta (1981) concluded that NGC~3077 is comprised
of ``a number of very hot young stars embedded in a smoother, cooler
component''.
This degree of recent star formation is confirmed by HST/NICMOS
observations made in the Paschen line by B\"{o}ker et al (1999).

Given its peculiar nature, NGC~3077 has been the focus of numerous
studies emphasizing its significance with respect to the formation
of dwarf galaxies and galaxy interactions.
One of the more important ``breakthrough'' discoveries that
greatly affected our understanding of the evolution of NGC~3077
was the detection of an HI bridge between M81 and this galaxy
(van der Hulst 1979).  More recent HI observations using the VLA
revealed the distribution of the gas at higher resolution, 
extending not only from M81 to NGC~3077, but also out to M82 (Yun et al. 1994).
Using these results, Yun et al. (1994) were able to produce
a numerical simulation of the dynamical evolutionary history of the M81-M82-N3077
system.
Although there had been several such simulations prior to Yun et al.
(Cottrell 1977, van der Hulst 1977, Killian 1978, Brouillet et al. 1991),
they were not as accurate; the fundamental difference between Yun et al. (1994)
and earlier studies was that the former were able to include an HI disk
surrounding NGC~3077.

At the periphery of NGC~3077, there is also a prominent stellar complex,
which was nicknamed the ``Garland'' by BBdT74.
The precise origin of this feature is uncertain, although it is plausibly
associated with the tidal interaction of NGC~3077 with M81
(Karachentsev, Karachentseva \& Borngen 1985: hereafter KKB85).
Some, however, suggest that this stellar complex is actually 
a progenitor of a dwarf galaxy (KKB85).
At an assumed distance modulus of $\sim  27.6~mag$, the brightest stars
at $B \sim 21~mag$ in the ``Garland" were thought to be blue
supergiants (KKB85).

As part of a long--term project to obtain distances to all the galaxies
in the local volume within $\sim$5 Mpc using the tip of the red giant
branch (TRGB) method, we have made V and I observations
of two regions in the halo of NGC~3077 using Wide Field Planetary Camera 2
(WFPC2) on board the Hubble Space Telescope (HST).
The TRGB method has been proven to be an excellent distance indicator
(da Costa \& Armandroff 1990; Lee et al. 1993; Sakai 1999),
especially for obtaining consistent distances to all morphological
types of galaxies, such as our case in which we are building a complete
volume--limited sample.
The NGC~3077 halo regions observed with HST were chosen to overlap 
with part of the ``Garland'' stellar concentration.
In this paper, we report in addition to the detection of the
red giant branch stars in NGC~3077, 
observations of resolved upper main sequence stars in the ``Garland''.
The observations and data reduction procedures are briefly described
in \S2.  We then discuss the detection of the RGB stars and the determination of
a distance
to NGC~3077 measured by the TRGB method (\S3), followed by a discussion 
of the upper main sequence stars observed in our frames (\S4).

\section{Observations}
Two positions in the halo region of NGC~3077 were chosen 
for our HST observations.  A digital sky survey image of NGC~3077,
is shown in Figure~1, on which the 
$HST$ Wide Field Planetary Camera 2 (WFPC2) footprints are 
superimposed, indicating the two regions (Fields I and II) observed.
They are located, respectively, $\sim$3' southwest and $\sim$5' south from 
the nucleus of the galaxy.
The Planetary Camera (PC) chip which covers the smallest area
is referred to here as Chip 1.
The three Wide Field (WF) chips cover the three larger fields and
are referred to as Chips 2, 3 and 4 respectively,
numbered counterclockwise from the PC.
A close-up $HST$ image of one of the chips, WF2 field of Field I, 
is shown in Figure~2, illustrating the high degree of resolution
into stars.
The Fields~I and II were observed on January 27, 1999 and October 23, 
1998 respectively.
Two exposures (400 seconds for Field I and 500 seconds for Field II)
were taken using two filters (F555W and F814W) at each position.
Cosmic rays on each image were cleaned before being combined to make a set
of F555W and F814W frames.  

The subsequent photometric analysis was done
using point spread function fitting packages DAOPHOT and ALLSTAR.
These programs use automatic star finding algorithms and then measure
stellar magnitudes by fitting a point spread function (PSF), constructed
from other uncrowded HST images (Stetson 1994).
We checked for a possible variation in the luminosity function as a function
of the position on each chip by examining the luminosity functions for
different parts of the chip.
For each frame, we find identical luminosity functions, 
confirming that there is no significant systematic offsets originating
from the adopted PSFs.

The F555W and F814W instrumental magnitudes were converted to the calibrated
Landolt (1992) system as follows.  (A detailed discussion is found in
Hill et al. 1998).  The instrumental magnitudes were first transformed
to the Holtzman et al. (1995) 0\Sec5 aperture magnitudes by determining
the aperture correction that need to be applied to the PSF magnitudes.
This was done by selecting 20--30 brighter, isolated stars on each frame.
Then all the stars were subtracted from the original image except for
these selected stars.  The aperture photometry was carried out for these
bright stars at 12 different radii ranging from 0\Sec15 to 0\Sec5.
The 0\Sec5 aperture magnitudes were determined by applying the growth
curve analysis provided by DAOGROW (Stetson 1990), which were then compared
with the corresponding PSF magnitudes to estimate the aperture corrections
for each chip and filter combination.
We use a different set of aperture corrections for the two fields.
Most of the values agree with each other within $2\sigma$,
however slight offsets between the corrections in the two fields are
most likely due to the PSFs not sampling the images in the exactly same way.
When images are co--added, the combined images are not exactly identical
to the original uncombined images;
that is, the precise positions of stars on the frames are slightly
different. Thus we should expect
some differences in the aperture corrections of the same chip in 
two fields.  

Finally, the 0\Sec5 aperture magnitudes were converted to the standard
system via the equation:
\begin{equation}
M = m + 2.5 \log t + {\mbox{C1}} + {\mbox{C2}} \times (V-I) + {\mbox{C3}} \times
(V-I)^2 + {\mbox{a.c.}},
\end{equation}
where $t$ is the exposure time, C1, C2 and C3 are constants and a.c. is the
aperture correction.
C1 is comprised of several terms including (1) the long--exposure WFPC2 magnitude
zero points, (2) the DAOPHOT/ALLSTAR magntidue zero point, (3) a correction
for multiplying the original image by 4 before converting it to integers
(in order to save the disk space), (4) a gain ratio term due to the difference
between the gain settings used for NGC~3077 and for the Holtzman et al. (1995) data 
(7 and 14 respectively), (5) a correction for the pixel area map which
was normalized differently from that of Holtzman et al., and
(6) an offset between long and short exposure times in the HST zero point
calibration.
C2 and C3 are color terms and are the same for all four chips.

In Table~1, we list the V and I magnitudes of selected bright reference stars in NGC~3077. 
The columns list the following:
(1) WFPC2 image root name; (2) chip number; (3) and (4) the (x,y) coordinates of the stars;
(5) and (6) right ascension and declination (J2000); (7) and (8) V magnitude and errors; 
and (9) and (10) I magnitude and errors.

\section{Detection of the Red Giant Branch Stars and the
Distance to NGC~3077}

In Figure~3, $(V-I) - I$ color magnitude diagrams for stars in
both observed halo fields of NGC~3077 are shown.
Because the exposure time was slightly shorter for the observations
of Field~I, the incompleteness magnitude is brighter for this field.
The CMD's can be characterized by their prominent red giant branches (at $(V-I) \sim 2$ mag),
and also by the upper main sequence (blue plume at $(V-I) \sim 0.0$ mag) 
stars especially evident in Field~II.
We will return to the discussion of the main sequence stars in \S4.

The top panels of Figure~4 show the I--band luminosity histograms for
all the stars in Fields I and II, whereas the bottom panels show
those restricted to the red giant stars defined as having colors 
between $1.0 < (V-I) < 3.5$ mag.
The color restriction eliminates obvious main--sequence stars,
and limits the sample of RGB stars.
The comparison of upper and lower histograms again suggests that
the Field~II stellar population is composed of a larger fraction of non-RGB stars.  

There is a jump, especially in Field~II, in the luminosity function at
$I \simeq 24.0$ mag. We identify this feature with the tip of the 
red giant branch (TRGB).
The TRGB marks the onset of core helium flash in old, low-mass stars.
These stars evolve up the red giant branch, but suddenly change their physical
characteristics upon the ignition of helium in their cores.
The magnitude at which this transformation takes place turns out to be
extremely insensitive to both metallicity and age (Iben \& Renzini 1983) in the I-band
($\sim$8200\AA), manifesting itself an abrupt discontinuity in the luminosity function
at M$_I \simeq -4$ mag.
The insensitivity of the TRGB magnitude in the I--band has been confirmed
both observationally and theoretically (da Costa \& Armandroff 1990, Lee,
et al. 1993, Salaris \& Cassisi 1997).
The empirical calibration (Lee et al. 1993) used consistently in this series,
is derived from the distances to
four Galactic globular clusters measured using the RR Lyrae distance scale
based on the theoretical horizontal branch model for $Y_{MS}=0.23$ of 
Lee, Demarque \& Zinn (1990) which corresponds to $M_V (\mbox{\small RR Lyrae}) 
= 0.57$ mag at [Fe/H] = $-1.5$.
The globular clusters used in the calibration span the metallicity range of
$-2.2 <$ [Fe/H] $< -0.7$ dex.

To measure the distance to NGC~3077 using the TRGB method, we select the
stars in WF Chips 2 and 4 of Field~I.  This is because (1) Field I
is less contaminated by upper main--sequence stars, and (2) WFC2 and 4
are centered on less--crowded regions.
The Gaussian--smoothed I--band luminosity function is shown in the
top panel of Figure~5.  A jump in the number count is clearly detected
at I $\sim$ 24.0 mag.
We apply an edge--detection filter, which is a modified version of
a Sobel kernel ([-1,0,+1]), to the luminosity function to
determine objectively the position of the TRGB following
$E(m) = \Phi(m_I + \sigma_{m_I}) - \Phi(m_I - \sigma_{m_I})$, where $\Phi(m)$ is
the number of stars per unit magnitude integral at magnitude $m$,
and $\sigma_m$ is the typical photometric error of stars at magnitude $m$.
For the details of the Sobel filter application, readers
are referred to the Appendix of Sakai, Madore \& Freedman (1996).
The results of the convolution are shown in the bottom panels of Figure~5.
The position of the TRGB is identified with the highest peak in the filter output
function.
For NGC~3077, we obtain \ITRGB $= 24.0 \pm 0.1$ mag.

We use the empirical calibration given by Lee et al. (1993) to determine
the distance to NGC~3077, which is expressed as $(m-M)_I =$ \ITRGB -
$M_{\mbox{\tiny bol}} + BC_I$.  The bolometric magnitude, $M_{\mbox{\tiny bol}}$, 
and the bolometric correction, $BC_I$, are given in terms of the metallicity of the RGB
stars:  $M_{\mbox{\tiny bol}} = -0.19[Fe/H]-3.81$ mag, and $BC_I = 0.881 - 0.243
(V-I)_{\mbox{\tiny TRGB}}$.
The metallicity, $[Fe/H]$, is expressed as a function of the $(V-I)$ color:
$[Fe/H] = -12.65 + 12.6 (V-I)_{-3.5} - 3.3 (V-I)^2_{-3.5}$, where $(V-I)_{-3.5}$
is measured at the absolute $I$ magnitude of $-3.5$mag.
For the halo stellar population in NGC~3077, we observe $(V-I)_{\mbox{\tiny TRGB}} =
2.25 \pm 0.6$ mag, and $(V-I)_{-3.5} = 2.0 \pm 0.5$ mag.
Substituting these values into above equations, we obtain
$BC_I - M_{\mbox{\tiny bol}} = 4.02 \pm 0.07$ mag, thus yielding 
$(m-M)_I = 28.02 \pm 0.14$ mag.
The foreground extinction in the direction of NGC~3077 is $A_B = 0.21$ mag
(Burstein \& Heiles 1982).
Using conversions of $A_V/E(V-I) = 2.45$ and $R_V = A_V/E(B-V) = 3.2$ (Dean,
Warren, \& Cousins 1978; Cardelli, Clayton, \& Mathis 1989),
we obtain $A_I = 0.09$ mag.
Thus, the dereddened distance modulus of NGC~3077 is $(m-M)_0 = 27.93 \pm 0.12_{\mbox
{\tiny random}} \pm 0.16_{\mbox{\tiny systematic}}$ mag,
which corresponds to a linear distance of 3.9 $\pm 0.2_{\mbox{\tiny r}}
\pm 0.4_{\mbox{\tiny s}}$ Mpc.
The random uncertainties include the $\pm0.1$ mag 
error in the edge--detection, and also the $\pm0.07$ mag uncertainty
in the TRGB absolute calibration due to the color spread of the RGB stars.
The two uncertainties are added in quadrature to yield the random error of $\pm 0.12$ mag.
The systematic uncertainties, on the other hand, come mainly from the TRGB
calibration ($\pm$0.15 mag) and the $HST$ photometry zero point ($\pm$0.05 mag).
The details of the nature of the systematic uncertainty have been discussed previously
in Sakai \& Madore (1999).

The distance to NGC~3077 we obtain in this paper compares very well with
other distance estimates to galaxies in the M81 group.  In Table~2, we tabulate
five galaxies whose distances have been estimated using either the TRGB method
or Cepheid variable stars.  Remarkably, all five distances agree within
1$\sigma$ of each other.
In the plane of the sky, the five galaxies have a projected transverse 
separation of approximately 120 kpc.
Along the line of sight, the depth of the M81 group formally 
extends for $\sim$350 kpc, but the uncertainty on the line-of-sight thickness is itself
about $\pm$250 kpc.
It is plausible that these five galaxies are all contained within a sphere of radius 60
kpc.

\section{Detection of Upper Main Sequence Stars in the Halo: Tidally Induced Star Formation?}

In Figure~3, we presented $I-(V-I)$ CMDs for the two fields in the halo of NGC~3077
observed using $HST$.  Field~II is positioned further away from the main body
of the galaxy compared to Field~I.  However, we also noted that there are a significant
number of upper main sequence stars in Field~II.  
In Figure~6, we show the distribution in the plane of the sky of all stars found 
using the DAOPHOT package.
Also overplotted by solid circles are the upper main sequence stars which
are defined on the CMD as those satisfying $I \leq 26.0$ mag and $-0.4 \leq (V-I) \leq 0.7$.
Two features can be observed here:  
(1) A concentration of blue stars is found around
RA(2000) =$10^h 03^m 10^s$ and DEC(2000)$=68^{\circ} $
42'~30'' which coincides with the outer
region of the main body of the galaxy;  
(2) A concentration of blue stars in the WF Chips 2  and 3 of Field~II.  
The distribution of the RGB stars
follows what is expected for the halos of galaxies, with its surface brightness
decreasing as a function of radial distance from the center of the galaxy.  
However, the upper MS stars
are distributed in a distinctively different fashion, characterized by concentrations
that are separated from the main body of NGC~3077.  
The region around RA=$10^h 03^m 35^s$ and DEC=$68^{\circ}$ 40' overlaps with the
edge of a stellar concentration nicknamed ``Garland''.
As mentioned in the \S1, the causal origin of this feature is uncertain, 
even though it is very likely associated with the previous tidal 
interaction of galaxies in the M81 group.

Although a careful analysis of the star formation history of young stars
in this region would require a better set of data, preferably with wider
spatial coverage and deeper photometry, we examine in this Section, the 
properties of these upper MS stars in the Garland region.

We begin by comparing our observed CMD with the Padua theoretical isochrones
(Bertelli et al. 1994).  Unfortunately, there is no published measurement
of the abundance of NGC~3077.  
However, Martin (1998) reports that the Pop~I metallicity of this galaxy
is likely larger than $1/3 Z_{\odot}$.
In her spectroscopic survey of nearby dwarf galaxies, she detected no
[OIII]$\lambda$4363 line in NGC~3077.
Based on the strength of [NII] $\lambda$6584, she suggests that the abundance
of NGC~3077 is at least $1/3 Z_{\odot}$, possibly much larger.
This Pop~I abundance is consistent with the Pop~II abundance in the sense that
the former is slightly larger than the latter
estimated above using the colors of the red giant branch stars 
([Fe/H] $\leq -0.6$,
which corresponds to $Z=0.005$).
However, we note here that the Pop~II metallicity inferred by the RGB stars
is rather high for a small dwarf galaxy such as NGC~3077.
The reddest RGB stars is comparable to that of, for e.g.,
47~Tuc, a Galactic globular cluster with the metallicity of [Fe/H]$\sim -0.7$.
This value is extremely high if NGC~3077 were to follow the luminosity-metallicity
relation for dwarf galaxies.  We will address this point later in this Section.
The Padua isochrone library provides the stellar evolution tracks for $Z=0.004$ and
$Z=0.008$.  We plot a set of isochrones for both of these abundances in Figure~7.
The number corresponding to each isochrone is the logarithm of age.
The main sequence stars span the ages from $\sim$30 Myr up to $\sim$125 Myr.

Another way of presenting the star formation timescale is 
the V-band luminosity function of upper main sequence stars which is shown in
Figure~8.  The main sequence turn off (MSTO) ages are indicated as well on this plot.
A simple artificial star test suggests that our V-band photometry is complete
down to $V \simeq 25.5$ mag, which is consistent with the luminosity function
turn-over at $25.5-26.0$ mag.
Again, these main sequence stars up to $V \simeq 25.5$ mag correspond to
those of MSTO ages up to $\sim$150 Myr.

The ``Garland'', where the upper main sequence stars are found,
was first noticed by BBdT74, and later described by Karachentsev et al. (1985) 
as the ``most peculiar'' galaxy among 40 dwarf galaxies in the M81 group in 
their earlier survey
(Karachentseva, Karachentsev \& Borngen 1985).
The origin of the Garland is mostly likely associated with the tidal interaction
of NGC~3077 with M81.
Several numerical analyses have been presented to provide a dynamical history
of the M81-M82-NGC~3077 group (Cottrell 1977, van der Hulst 1977, Killian 1978,
Brouillet et al. 1991, Yun 1997). The most recent one by Yun (1997) was extremely
successful in not only reproducing the spatial distribution of HI, but also
its velocity field.  Unlike the previous numerical simulations, Yun (1997) 
was also able to produce the North Tidal Bridge which runs between M82 and
NGC~3077 by giving NGC~3077 its own disk (of nearly $10^9$M$_{\odot}$) initially.
The disk was then disrupted as the galaxy interacted with M81.  
Yun's model predicts that the nearest approach of NGC~3077 and M81 was 
280 Myr ago; for M82, the time since the pericenter is 220 Myr,
which agrees very well with the estimated age of ongoing starbursts in M82.
This strongly supports the idea that was first proposed by Toomre \& Toomre (1972)
such tidal disruptions could ``stoke the furnace'', enhancing starburst
or AGN activity.

Our HST data suggest that the stars were formed in the halo of NGC~3077
during past 150 Myrs, well after the calculated tidal encounter of this galaxy with M81.
This suggests that the tidal interaction possibly triggered star formation in
the halo region, perhaps from
a condensed interstellar medium that was stripped from the main body of the galaxy
as a result of the interaction.
A composite HI-optical map shown in Figure~1 of Walter \& Heithausen (1999)
displays very well that there is an offset between the center of HI concentration
and the nucleus of NGC~3077; with the HI concentrated to the south of the galaxy.
Our HST fields were placed in the south - southwestern region of the halo,
coinciding with the HI concentration.
This, in conjunction with the ages of the upper main-sequence stars, 
suggests that the tidal interaction is indeed responsible for the
recent star formation in the halo of NGC~3077.

Sakai \& Madore (1999) presented the HST/WFPC2 observations of halo stars
in M82.  We reported the detection of intermediate-age asymptotic giant branch
(AGB) stars, which again coincided with the HI concentrations in the M82 halo.
Because of the quality of the photometry which was not superior, and
also uncertainties due to blending, we concluded that the detection of
AGB stars was not 100\% certain.
Here in the case of NGC~3077, we observe a similar situation.
However, in addition to some AGB stars, we also observe blue upper MS stars,
indicative of very recent star formation.

It is easy to argue that the AGB stars are likely blends of two or more fainter
red stars.  However, in order to make a bright blue upper main sequence star,
one needs two fainter blue stars; that is, the existence of upper main sequence
stars, which signify a more recent star formation, cannot be dismissed.
For the case of Garland in NGC~3077, it is conceivable that the recent
star formation ($<150$Myr) was triggered as a direct result of the tidal interaction
of the galaxy with M81 (and M82).
However, the theoretical dynamical model cannot constrain the interaction timescale
by better than a factor of two, suggesting another scenario in which
these recently-formed upper MS stars were simply stripped from the main body of
NGC~3077, and not necessarily formed in the halo region.
Had the photometric quality of the stars been better (i.e. longer exposure times),
we might have been able to separate the MS stars from the blue He-burning (HeB) stars.
Using the HeB stars, one can trace the star formation history back to $\sim$600 Myr,
which was the technique used by Dohm-Palmer et al. (1997) to study the recent
star formation history in Sextans~A.
If we assume that these halo MS stars were indeed tidally-induced stars,
we are then able to put a lower limit of the timescale at which the tidal
interactions occurred.

As pointed out earlier, the Pop~II metallicity of some of the halo RGB stars in NGC~3077
is as high as [Fe/H]$ \sim -0.7$ which is uncharacteristic for a dwarf galaxy
of this size.
We note, however, that even though the current size and luminosity of NGC~3077
might suggest that this galaxy is similar to SMC, the numerical simulations (Yun 1997)
indicated that NGC~3077 had its own disk before interacting with M81.
The total mass given to NGC~3077 in the best three-body numerical model
was $2 \times 10^{10}$ M$_{\odot}$.
This mass is a couple of orders larger than that of a typical dwarf galaxy,
inferring that fairly metal-rich Population~II RGB stars could have populated
NGC~3077, which were then redistributed during the interaction with M81,
and now occupy the halo.
A slightly different scenario would be that the more metal-rich RGB stars
were originally part of M81 which were pulled apart during their collision,
and eventually settled down in the halo of NGC~3077.
It is impossible to determine the exact nature or origin of these
halo stars, but on the other hand, we know from the deep HI maps of the
M81 group that these galaxies have undergone a major interaction with each other;
we cannot by any means regard NGC~3077 as an isolated dwarf galaxy.
The metal-rich RGB stars in this small galaxy may be 
direct evidence N3077 was originally much larger spiral galaxy
before the interaction.

Finally, another question of interest is
whether the Garland is simply an interstellar medium and stars stripped from the main
body of NGC~3077, or is an independent dwarf galaxy.
This is beyond the scope of this paper, requiring the photometric observation of the entire
Garland region, and also mapping its velocity field.

\medskip
This work was funded by 
and also by NASA LTSA program, NAS7-1260 to SS.
BFM was supported in part by the NASA/IPAC Extragalactic Database.

\newpage

{\bf Figure Captions}

{\bf Figure 1:}  A digital sky survey image of NGC~3077, with the $HST$ WFPC2 footprints
superimposed, indicating the two regions (Fields I and II) observed.

{\bf Figure 2:} A close-up $HST$ image of one of the chips, WF2 field of Field I, 
illustrating the high degree of resolution into stars.

{\bf Figure 3:}  $(V-I) - I$ color magnitude diagrams for stars in
both observed halo fields of NGC~3077.

{\bf Figure 4:} {\it (top): } I--band luminosity histograms for
all the stars in Fields I and II.  {\it (bottom):} restricted to the
red giant stars with colors between $1.0 < (V-I) < 3.5$ mag.
The color restriction should eliminate obvious main--sequence stars,
and limit the sample to RGB stars.

{\bf Figure 5:} {\it (top): } The Gaussian--smoothed I--band luminosity function.
A jump in the number count is clearly observed at I $\sim$ 24.0 mag.
{\it (bottom): } output of an edge-detection filter applied to the luminosity function.
The position of the TRGB is indicated by the highest peak in the output function.

{\bf Figure 6:} The distribution in the plane of the sky of all stars found 
using the DAOPHOT package.
Also overplotted by solid circles are the upper main sequence stars which
are defined on the CMD as those satisfying $I \leq 26.0$mag and $-0.4 \leq (V-I) \leq 0.7$ mag.
The cross represents the position of 
the nucleus of NGC~3077, which is at RA = 10$^h$ 03$^m$ 20.3$^s$ and DEC = 68$^{\circ}$ 44$^{\prime}$ 
02$^{\prime\prime}$.

{\bf Figure 7:} $(V-I)-V$ CMDs for all the stars found in our HST/WFPC2 fields.
Superimposed are the Padua isochrones for $Z = 0.004$ {\it (left)} and for
$Z = 0.008${\it (right)}.  
The number corresponding to each isochrone is the logarithm of age.

{\bf Figure 8:} The V-band luminosity function of upper main sequence stars.
The main sequence turn off ages are indicated as well on this plot.
A simple artificial star test suggests that our V-band photometry is complete
down to $V \simeq 25.5$ mag, which can also be seen in the luminosity function
turn-over at $25.5-26.0$ mag.

\medskip
\medskip

\newpage

\setcounter{table}{0}
\begin{deluxetable}{cccccccccc}

\tablecaption{Positions and Magnitudes of Bright Stars in NGC 3077}
\tablewidth{0pc}
\tablecolumns{10}
\tablehead{
\colhead{Frame} &
\colhead{Chip} & 
\colhead{X} &
\colhead{Y} &
\colhead{RA } &
\colhead{DEC} &
\colhead{$V$} &
\colhead{$\sigma_V$} &
\colhead{$I$} &
\colhead{$\sigma_I$} \cr
\colhead{Rootname} &
\colhead{} &
\colhead{} &
\colhead{} &
\colhead{(J2000)} &
\colhead{(J2000)} &
\colhead{(mag)} &
\colhead{(mag)} &
\colhead{(mag)} &
\colhead{(mag)}
}
\startdata 
U3NK5101R & 2 & 692.1 & 334.1 & 10:02:49.05 & 68:41:23.1 & 22.93 &  0.04 & 22.54 &  0.03 \cr
  &    2 & 674.1 & 266.1 & 10:02:49.77 & 68:41:17.4 & 23.63 &  0.05 & 21.67 &  0.03 \cr
  &    2 & 722.2 & 627.3 & 10:02:46.79 & 68:41:49.6 & 23.45 &  0.05 & 23.09 &  0.05 \cr
  &    2 & 654.9 & 408.4 & 10:02:49.24 & 68:41:31.3 & 23.61 &  0.05 & 23.27 &  0.06 \cr
  &    2 & 179.2 & 439.0 & 10:02:57.24 & 68:41:49.7 & 24.00 &  0.05 & 23.79 &  0.07 \cr
  &    2 & 172.5 & 489.6 & 10:02:57.05 & 68:41:54.7 & 23.95 &  0.08 & 23.71 &  0.07 \cr
  &    3 & 766.1 & 193.7 & 10:03:00.17 & 68:42:28.7 & 21.71 &  0.08 & 20.14 &  0.04   \cr
  &    3 & 368.8 & 714.0 & 10:03:11.45 & 68:42:08.1 & 22.43 &  0.07 & 22.42 &  0.06   \cr
  &    3 & 544.8 & 603.1 & 10:03:08.51 & 68:42:21.1 & 23.15 &  0.10 & 22.95 &  0.07   \cr
  &    3 & 171.2 & 674.8 & 10:03:11.93 & 68:41:48.3 & 23.34 &  0.07 & 23.31 &  0.08   \cr
  &    3 & 480.5 & 465.5 & 10:03:06.52 & 68:42:10.7 & 23.99 &  0.09 & 23.47 &  0.08   \cr
  &    3 & 773.5 & 158.1 & 10:02:59.52 & 68:42:28.2 & 23.17 &  0.08 & 22.61 &  0.08   \cr
  &    3 & 554.0 & 420.0 & 10:03:05.30 & 68:42:16.1 & 24.06 &  0.08 & 23.21 &  0.09   \cr
  &    4 & 279.3 & 347.2 & 10:03:07.70 & 68:40:54.2 & 19.18 &  0.05 & 17.44 &  0.04 \cr
  &    4 &  59.7 & 367.3 & 10:03:04.03 & 68:40:45.4 & 23.05 &  0.07 & 22.45 &  0.05 \cr
  &    4 & 346.2 & 245.8 & 10:03:08.28 & 68:41:05.8 & 22.07 &  0.06 & 20.87 &  0.05 \cr
  &    4 & 180.2 & 441.8 & 10:03:06.53 & 68:40:42.1 & 23.06 &  0.06 & 22.75 &  0.05 \cr
U3NK0105R & 2 & 284.4 & 352.2 &  10:03:28.70 & 68:40:28.6 &  23.90 &  0.04 & 22.51 &   0.03  \cr
 &      2 & 237.6  & 767.1   & 10:03:36.27 & 68:40:31.6 & 23.32  &  0.03  & 21.78  &  0.03  \cr
 &      2 & 358.7  & 724.4   & 10:03:35.11 & 68:40:42.6 & 23.63  &  0.05  & 22.02  &  0.03  \cr
 &      2 & 190.2  & 762.3   & 10:03:36.33 & 68:40:26.9 & 22.97  &  0.03  & 21.44  &  0.04  \cr
 &      2 & 561.4  & 196.6   & 10:03:25.00 & 68:40:52.8 & 24.32  &  0.07  & 22.99  &  0.04  \tablebreak
 &      2 &  97.2  & 386.3   & 10:03:29.93 & 68:40:10.9 & 22.63  &  0.02  & 22.57  &  0.04  \cr
 &      2 & 472.0  & 603.1   & 10:03:32.57 & 68:40:51.4 & 24.01  &  0.04  & 22.61  &  0.04  \cr
 &      2 & 205.7  & 729.6   & 10:03:35.71 & 68:40:27.8 & 23.15  &  0.03  & 22.93  &  0.05  \cr
 &      2 &  82.0  & 405.6   & 10:03:30.33 & 68:40:09.8 & 24.19  &  0.05  & 22.89  &  0.05  \cr
 &      2 & 341.6  & 612.4   & 10:03:33.17 & 68:40:38.9 & 23.44  &  0.04  & 23.57  &  0.06  \cr
 &      3 & 204.1  & 217.9   & 10:03:27.25 & 68:39:46.1 &  22.80 &  0.05  & 20.77  &  0.04  \cr
 &      3 & 556.1  & 402.0   & 10:03:34.15 & 68:39:34.3 &  21.65 &  0.05  & 21.58  &  0.05  \cr
 &      3 & 480.4  & 269.2   & 10:03:32.37 & 68:39:46.0 &  22.93 &  0.05  & 20.78  &  0.05  \cr
 &      3 & 661.2  & 305.5   & 10:03:35.72 & 68:39:45.5 &  23.23 &  0.06  & 21.53  &  0.05  \cr
 &      3 & 369.5  & 467.0   & 10:03:31.01 & 68:39:24.7 &  22.79 &  0.06  & 21.62  &  0.07  \cr
 &      3 & 510.5  & 336.1   & 10:03:33.12 & 68:39:39.9 &  23.78 &  0.07  & 22.21  &  0.07  \cr
 &      3 & 581.9  & 601.1   & 10:03:35.25 & 68:39:15.3 &  23.18 &  0.06  & 23.05  &  0.09  \cr
 &      4 &  98.2  &  56.1   & 10:03:23.61 & 68:39:54.3 &  22.96 &  0.07  & 22.53  &  0.07  \cr
 &      4 & 141.0  & 721.1   & 10:03:11.85 & 68:39:39.1 &  23.29 &  0.07  & 22.74  &  0.08  \cr
 &      4 & 348.0  & 447.6   & 10:03:17.37 & 68:39:23.5 &  23.41 &  0.07  & 22.95  &  0.08  \cr
 &      4 & 780.1  & 336.9   & 10:03:20.66 & 68:38:43.1 &  23.67 &  0.08  & 23.05  &  0.09  \cr
\enddata
\end{deluxetable}

\setcounter{table}{1}
\begin{deluxetable}{cccccc}
\def\Deg{\hbox{${}^\circ$\llap{.}}}
\def\Min{\hbox{${}^{\prime}$\llap{.}}}
\def\Sec{\hbox{${}^{\prime\prime}$\llap{.}}}

\tablecaption{Distances to Selected Galaxies in the M81 Group}
\tablewidth{0pc}
\tablecolumns{6}
\tablehead{
\colhead{Galaxy} &
\colhead{RA (J2000)} &
\colhead{DEC (J2000)} &
\colhead{Distance Modulus} &
\colhead{Method} &
\colhead{Reference} \cr
\colhead{} &
\colhead{} &
\colhead{} &
\colhead{(mag)} &
\colhead{} &
\colhead{} \cr
}
\startdata 
M81      & 09 55 33.2 & +69 03 55  & 27.8 $\pm$ 0.20  & Cepheids & Freedman et al. (1994) \cr
M82      & 09 55 52.2 & +69 40 48  & 27.95 $\pm$ 0.14 & TRGB & Sakai \& Madore (1999) \cr
NGC~3077 & 10 03 19.2 & +68 43 59  & 27.93 $\pm$ 0.14 & TRGB & this paper \cr
BK5N     & 10 04 41.1 & +68 15 22  & 27.9  $\pm$ 0.15 & TRGB & Caldwell et al. (1998) \cr	
F8D1     & 09 44 47.1 & +67 26 19  & 28.0 $\pm$ 0.10  & TRGB & Caldwell et al. (1998) \cr
\enddata
\end{deluxetable}

\end{document}